\documentclass[a4paper, 12pt]{article}
\usepackage[english]{babel}
\usepackage[utf8]{inputenc}
\usepackage{amsmath}
\usepackage{indentfirst}
\usepackage{graphicx}
\usepackage{comment}
\usepackage{multicol,lipsum}
\usepackage{amsfonts}

\usepackage[margin=1.0in]{geometry}

\begin{document}

\begin{titlepage}
    \begin{center}
    
        \textbf{\Huge{Multiple Angles of Arrival Estimation using Neural Networks}} \\
        \vspace{15pt}
        \large{Ph.D. Qualifying Exam Report} \\
        
        \vspace{15pt}
        \vspace{95pt}
        \textbf{\LARGE{Jianyuan Yu}}\\
        \vspace{10pt}
        \large{Bradley Department of Electrical and Computer Engineering, \\Virginia Tech, USA}
        \vspace{3,5cm} \\
        \large{Committee:} \\
        \large{Dr. Jeffrey H. Reed} \\
        \large{Dr. R. Michael Buehrer}\\
        \large{Dr. Harpreet S. Dhillon}\\
    \end{center}

    \vspace{1cm}
    
    \begin{center}
        \vspace{\fill}
        \large{May 20, 2019} \\
        
    \end{center}
\end{titlepage}

\newpage
\tableofcontents
\thispagestyle{empty}

\newpage
\pagenumbering{arabic}
\section{Introduction of Neural Network}
\subsection{Machine learning and artificial intelligence}

In the recent decade, machine learning (ML) has assisted from self-driving cars, practical speech recognition, effective web search, to a vastly improved understanding of the human genome. It is defined as a scientific study of algorithms and statistical models that computer systems use to effectively perform a specific task without using explicit instructions, relying on patterns and inference instead. The point contributes to its diversity is that machine learning builds a mathematical model based on sample data, known as training data, in order to make predictions or decisions without being explicitly programmed to perform the task.

Artificial Intelligence (AI) refers to the intelligence demonstrated by machines, in contrast to the natural intelligence displayed by humans and animals. In general, artificial intelligence aims at building a robot in a way that works like a human being, which can learn, make mistakes, compensate for your mistake, and gaining an experience that leads to learning. Artificial intelligence is mainly cover the fields such as image processing, signal processing, natural language processing, databases and etc.

\begin{figure}[h]
\centering
    \includegraphics[width=.5\linewidth]{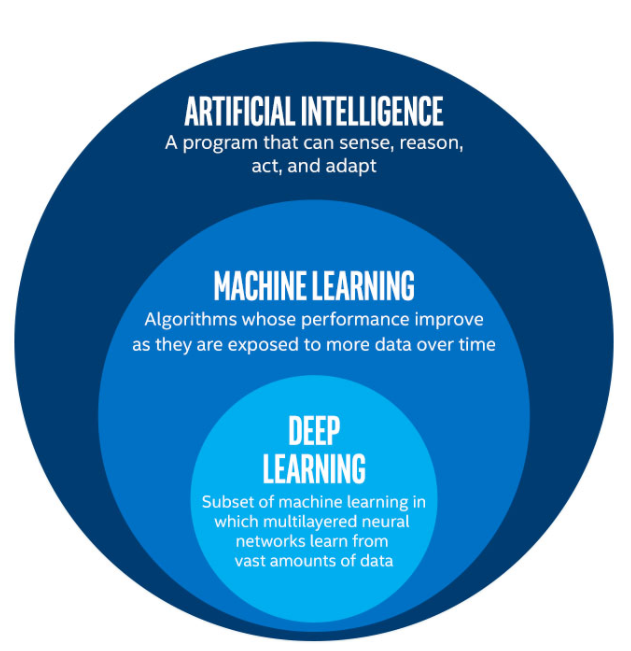}
    \caption{Machine Learning and Artificial Intelligence}
    \label{fig:MLvsAI}
\end{figure}

Machine learning is considered as a vital subset of artificial intelligence. Fig.~\ref{fig:MLvsAI} well demonstrates the difference between them. Artificial intelligence is a broader concept than machine learning, which addresses the use of computers to mimic the cognitive functions of humans, when machines carry out tasks based on algorithms in an intelligent manner. While machine learning focuses on the ability of machines to receive a set of data and learn for themselves, changing algorithms as they learn more about the information they are processing.

\subsection{Supervised learning, unsupervised learning and reinforcement learning}

The machine learning is further categorized into supervised learning, unsupervised learning and reinforcement learning. The supervised learning means the task of learning a function that maps an input to an output based on example input-output pairs. It infers a function from labeled training data consisting of a set of training examples, and the function can be used for mapping new examples. Here it take input variables $x$ and an output variable $y$ and an algorithm is used to learn the \textit{mapping function} from the input to the output $y = \tilde{f}(x)$. The goal is to approximate the mapping function  $\tilde{f}(.)$ so well that when you have new input data $x$ that you can predict the output variables $\hat{y}$ for that data. Depend on the output is categorized or not, the supervised learning problem could be further divided into classification and regression. The former predicts the categorical response value where the data can be separated into specific classes, while the later mostly predict the continuous-response value. Traditional method includes linear regression, random forest and Support Vector Machines (SVM), etc.

As a contrast, unsupervised learning only provides input data $x$ and no corresponding output variables or labels. It is named because there are no correct answers and there is no teacher, unlike supervised learning above. The goal of unsupervised learning is to model the underlying structure or distribution in the data in order to learn more about the data. Unsupervised learning problems can be further grouped into clustering and association problems. A clustering problem aims at discovering the inherent groupings in the data, such as grouping customers by purchasing behavior, while association explores rules that describe large portions of your data, such as people that buy $x_1$ also tend to buy $x_2$. Algorithms such as hierarchical clustering, $k$-means and apriori algorithm.

\begin{figure}[h]
\centering
    \includegraphics[width=.6\linewidth]{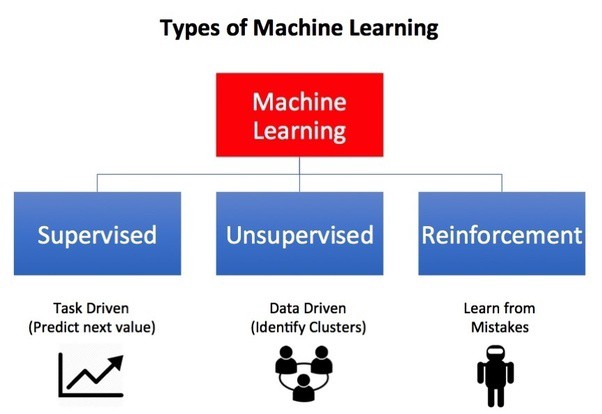}
    \caption{supervised, unsupervised and reinforcement learning}
    \label{fig:learning}
\end{figure}

Reinforcement learning concerned with how learning agents ought to take actions in an environment so as to maximize some notion of cumulative reward. Instead of labeled input or output pairs, actions by agent and reward from the environment is needed. Also, sub-optimal actions need not be explicitly corrected, while the focus is finding a balance between exploration and exploitation. The environment is typically formulated as a Markov decision process (MDP), as many reinforcement learning algorithms for this context utilize dynamic programming techniques. However, reinforcement learning algorithms differ from classical dynamic programming methods like MDP, is that it does not assume knowledge of an exact mathematical model of the MDP and they target larger scale MDPs. Besides, reinforcement learning can be done via online learning, which means it does not need a distinct training phase and testing phase, 
Fig.~\ref{fig:learning} well demonstrates the difference between them. To sum up, supervised learning training with labels, while unsupervised learning needs to learn the label by itself, reinforcement learning takes executive actions and immediate feedback from the environment as a way of learning.

\subsection{Neural network}

\begin{figure}[h]
\centering
    \includegraphics[width=.6\linewidth]{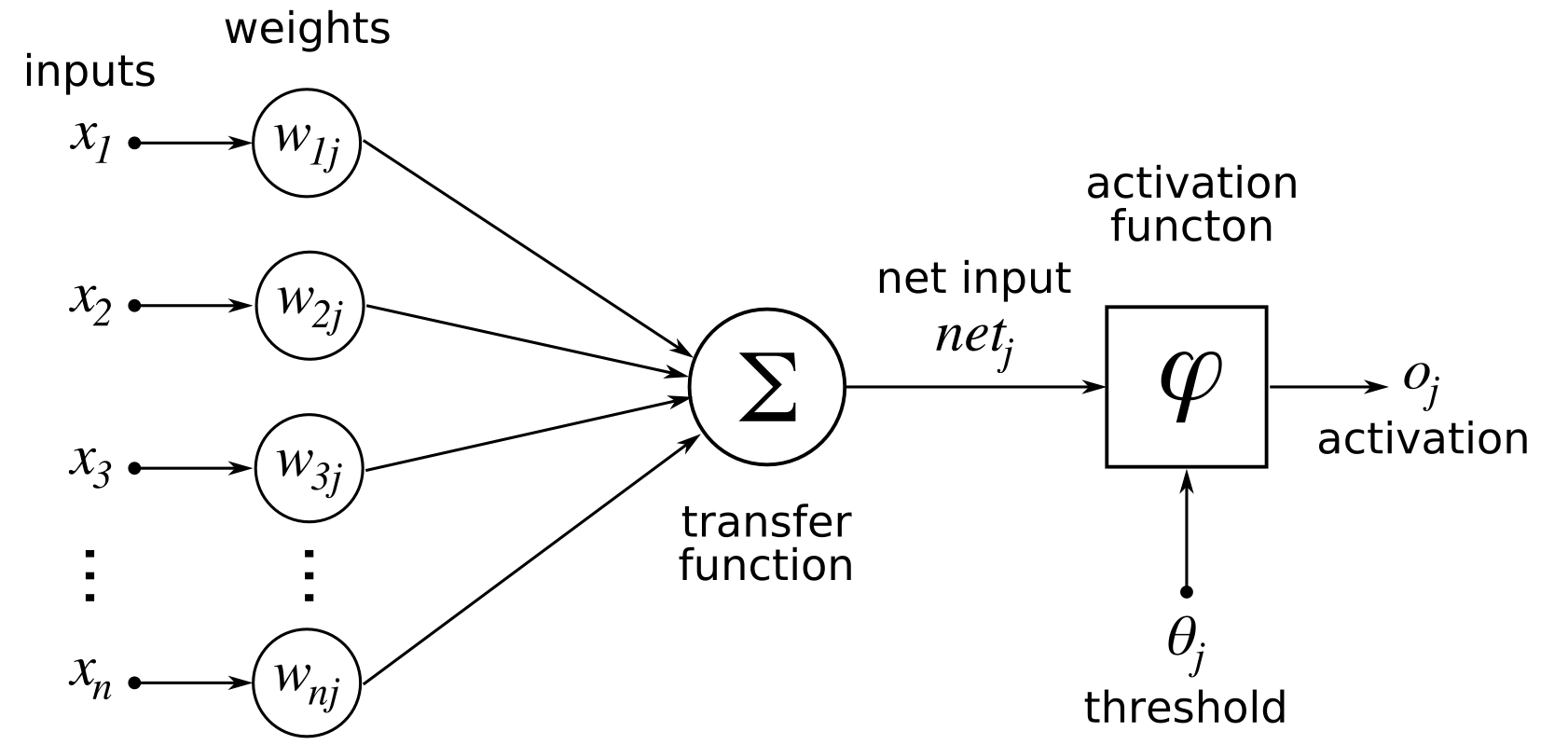}
    \caption{neuron}
    \label{fig:neuron}
\end{figure}
The most notable power of recent machine learning algorithm is based on the neural network structure, which refers to a network or circuit of artificial neurons. The network consists of an input layer, several hidden layers and an output layer. Each layer consist of a number of neurons, which is a basic weight-sum operation unit with based and activation function $f(.)$ to improve its performance.
\begin{equation}
    y_j = f(\sum_{i=1}^{N}x_i w_i)
\end{equation}

The connecting line denotes the weight from the previous neuron to the next neurons. These weights would be repeatedly updated by backpropagation. The update will be terminated once the training accuracy meets satisfactory level, weight value becomes stable or some other conditions are met. Fig.~\ref{fig:neuron} show the function of a neuron, which is the most basic unit of a neural network, it commonly represent a weight-sum operation of the inputs, with biased and activation function to improve its performance. These value of weights and baise would be updated during the training by \textit{back-propagadation}, while activation function is heuristic and chose depends on the problem.

\begin{equation}
    \textbf{w} = \textbf{w} - \partial L/ \partial \textbf{w}
\end{equation}

\subsubsection{Various types of neural networks}

\begin{figure}[h]
\centering
    \includegraphics[width=.4\linewidth]{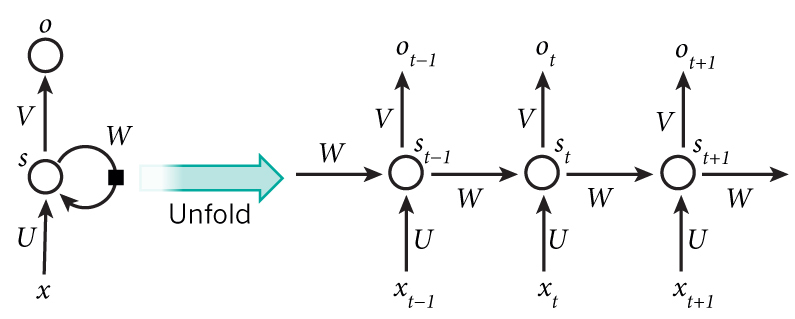}
    \caption{RNN}
    \label{fig:RNN}
\end{figure}

\begin{figure}[h]
\centering
    \includegraphics[width=.8\linewidth]{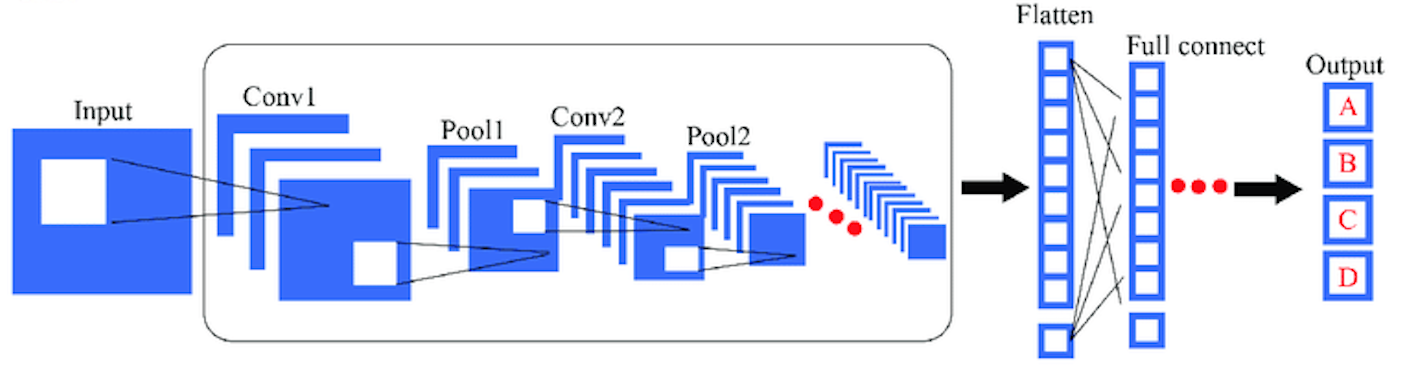}
    \caption{CNN}
    \label{fig:CNN}
\end{figure}

The most basic neural network consists of an input layer, a hidden layer and an output layer. Such a network is named \textit{shallow network}. While it is named \textit{deep network} if more than one hidden layer \cite{bianchini2014complexity}. The larger network generally increases the complexity and cost more training time, however, the accelerated computing of GPU makes it bearable nowadays. ANN is widely used for classification and function approximation. Besides, Neural Network has evolved from forwarding artificial neural network to Recurrent Neural Networks (RNN), Convolution Neural Network(CNN) and many other variations. Various type of neural networks has been studied, mainly depend on how neurons are connected, and we briefly introduce these two here.

RNNs is to make use of sequential information, rather than all inputs (and outputs) are independent of each other in ANN.  As shown in Fig.~\ref{fig:RNN}, it gets two flows, output from previous layerm and time propagation .RNNs are called recurrent because they perform the same task for every element of a sequence, with the output being depended on the previous computations. Another way to think about RNNs is that they have a memory which captures information about what has been calculated so far. This makes them applicable to tasks such as connected handwriting recognition or speech recognition. While CNN may fail at long-term dependency problem, and Long Short-Term Memory (LSTM) is proposed to fix it, with additional input gate, an output gate and a forget gate within a unit.

On the other hand, CNN is able to successfully capture the spatial and temporal dependencies in an image through the application of relevant filters as shown in Fig.~\ref{fig:CNN}. The architecture performs a better fitting to the image dataset due to the reduction in the number of parameters involved and reusability of weights. A convolution layer is added to extract the high-level features such as edges from the input image, and usually another pooling layer is added to get the average or max value among several inputs to reduce the computing.


\subsubsection{Activation function}
\begin{figure}[h]
\centering
    \includegraphics[width=.6\linewidth]{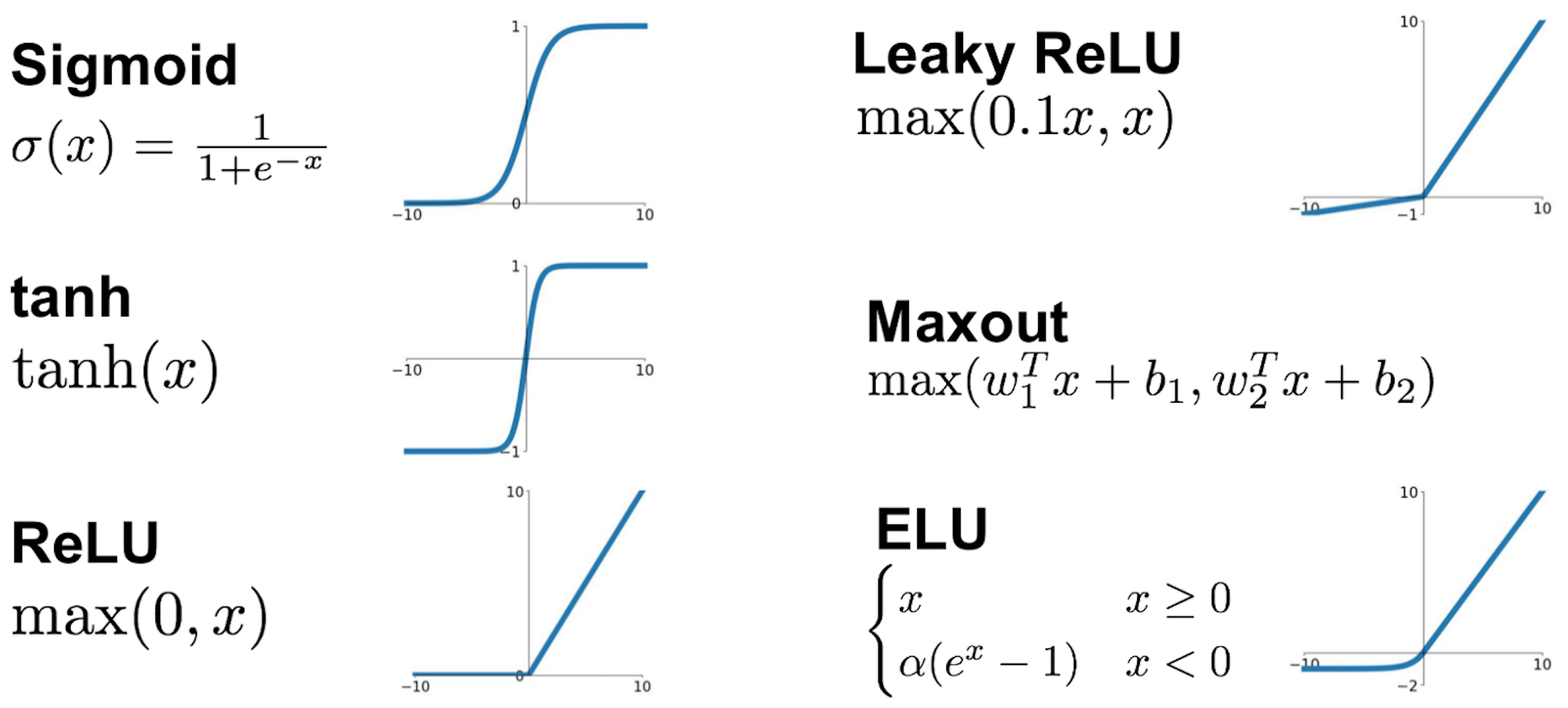}
    \caption{activation function}
    \label{fig:activiation}
\end{figure}

There are many detailed options and design within a neural network after set ANN or CNN. Here we take the activation function for example.  activation function, also known as transfer function is used to map input nodes to output nodes in a certain fashion. As shown in Fig.~\ref{fig:activiation}, several major activation function is list here. \textit{sigmoid} is commonly set as default in the neural network, and \textit{tanh} can extend the input range, \textit{Relu} remove the negative part, and Leaky Relu suppresses the negative part. There have not been guarantee which one is best among them, depend on different problems.

The radial activation function is also listed here, although it is expired in the modern library. However, it performs well in our problems \cite{el1997performance}.

\subsubsection{Applications}
Due to its model-free feature, and machine learning become popular in many other engineering fields, as we briefly sum up here.

\begin{itemize}
    \item \textbf{Signal processing}:
    Auto encoding has shown its advantage as unsupervised learning in the work \cite{o2017introduction}, and parameter estimation with neural network \cite{fehske2005new} can get precise results. It also covers topic like sequential learning, sequential decision methods, information-theoretic learning, graphical and kernel models and so on. 
    \item \textbf{Languange processing}: RNN is widely used in human language detection both in speech and audio, keyword extraction or matching, and smart chatbox \cite{manikopoulos2002network}.
    \item \textbf{Image processing}: Machine Learning unarguably boost the Computer Vision (CV) and Virtual Vision (VR) technology, it makes image classification, text recognization, human detection possible, it also benefits area like social networks, games, smart grid.
    \item \textbf{Communication systems}: machine learning could be the potential solution for various wireless communication problem  \cite{alsheikh2014machine}\cite{clancy2007applications} . It is shown Deep Q Network could help cognitive radio to share the spectrum with the interaction with the environment ~\cite{wang2018deep}. With the development of dense network and increasing Internet of Things(IoT) device, unsupervised learning could benefit the device-to-device users clustering \cite{fehske2005new}, grid user classification, anomaly or intrusion and so on. Meanwhile, supervised learning shows its advantage at MIMO channel estimation and detection, user location \cite{nerguizian2006geolocation}, and direction of arrival estimation. 
\end{itemize}


\newpage

\section{AoA Estimation by Neural Network}

\subsection{Related work of AoA estimation}

The angle of Arrival (AoA), also called the direction of arrival (DoA), is defined as the elevation and azimuth angle of incoming signals. Azimuth is the angle we are interested in this work. It usually requires antenna arrays, and apply widely in the field like localization, tracking, gesture recognition. The topic has a long history and gets a lot of techniques, and they can be divided into two categories, beamforming and subspace method. Moreover, Beamforming approaches \cite{van2004optimum} includes Bartlett method, Minimum Variance Distortion less Response (MVDR), and Linear prediction, while subspace-based approaches mainly refer to MUltiple SIgnal Classification (MUSIC) and its variants and  Estimation of Signal Parameters via Rotational In-variance Techniques (ESPRIT) \cite{li1991angle}, following work such as root-MUSIC  \cite{barabell1983improving} also could highly reduce the computing, and ESPRIT get smaller computing while constraint to larger asymptotic variance and apply for coherent signals. Meanwhile, some other approaches \cite{gezici2008survey} include k nearest neighbor (KNN) \cite{roshanaei2009dynamic}, Support-Vector Machine (SVM) \cite{pastorino2005smart} \cite{randazzo2007direction}, and some work skip estimating the DoA and directly estimate the position \cite{nerguizian2006geolocation} \cite{fang2008indoor}. However, these neural network has not extended to modern training techniques, neither tried the BNN, which would be our contributions here.

\subsubsection{The signal model of ULA}

\begin{figure}[h]
\centering
    \includegraphics[width=.4\linewidth]{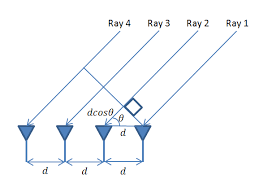}
    \caption{Uniform Linear Array}
    \label{fig:ULA}
\end{figure}

Far-field or plane wave assumption is commonly used in our problem, which refers to wavelength much smaller than the source to antenna distance. The signal model \cite{van2004optimum} of ULA follow the equation
\begin{equation}
x_m(t) = \sum_{n=1}^N s_n(t)e^{j2\pi \tau_{n}(m-1)}+n_m(t)
\end{equation}
for $M$ antennas, where $\tau_{n} = d cos(\theta_n)/\lambda$, $s_n(t)$ is $n$-th source signals which could simple be $ sin(2 \pi f_c t)$ or digital signals, $\tau_n$ phase shift per antenna, $N$ the number of sources and $n_m(t)$ noise terms. In matrix form it is represented as

\begin{equation}
X = AS + N
\end{equation}, where 
\begin{equation}
A = [a(\theta_1), a(\theta_2),...a(\theta_N)]
\end{equation}, and steering vector $a(\theta)$ is
\begin{equation}
a(\theta_i) = [1, e^{j2\pi\tau(\theta)},e^{j2\pi\tau(\theta)2},..., e^{j2\pi\tau(\theta)(M-1)}]^T
\end{equation}. Hence it also can be represent as 

\begin{equation}
     \begin{bmatrix} x_1(t)^T \\ x_2(t)^T \\ \vdots \\ x_M(t)^T \end{bmatrix}
 =
 [a(\theta_1), a(\theta_2),...a(\theta_N)]
    \begin{bmatrix} s_1(t)^T \\ s_2(t)^T \\ \vdots \\ s_N(t)^T \end{bmatrix} +
 \begin{bmatrix} s_1(t)^T \\ s_2(t)^T \\ \vdots \\ s_M(t)^T \end{bmatrix}
\end{equation}

There is also a close-from signal model for many other antennas geometry, such as circular arrays. Neural Network could also be potential to estimate since it works regardless of the signal model. Besides, We assume the ideal channel, without decay or delay at the receiver side, only white Gaussian noise is considered here.



\subsubsection{MUSIC}

As shown before, beamforming is a way of shaping received signals, it can be used for estimating AoA, also be used for directional communications. While Subspace-based approaches are specially designed for parameter (i.e.AoA) estimation using received signals, rather than used for extracting signals arrived from a certain direction. Subspace-based approaches decompose the received signals into \textit{signal subspace} and \textit{noise subspace}, so as to leverage special properties of these subspaces for estimating AoA.

The key idea of MUSIC  \cite{mathews1994eigenstructure} is to find a vector $q$ and a vector function $f(\theta)$, such that $q^Hf(\theta)=0$ if and only if $\theta = \theta_i$. The $p(\theta)$ is calculated by
\begin{equation}
    p(\theta)= \frac{1}{||q ^ H f(\theta)||^2}
\end{equation}, and the peaks of it indicates AoA since $q^H f(\theta)$ reach $0$. Moreover, MUSIC gives a way to find a pair of $a$ and $f(\theta)$. Recall that The signals from antenna array is $ X= AS+N$, and Covariance matrix of the signals is
\begin{equation}
\begin{split}
    R_{XX} & = E[XX^H] \\
    &= E[ASS^HA^H] + E[NN^H], \\
    & = A E[SS^H]A^H + \delta^2 I \\
    & = A R_{SS} A^H + \delta^2I,
\end{split}
\end{equation}
Here the former is signal terms and later is noise terms. Notice $A$ has  full column rank, and  signal term  $R_{SS}$ get rank $N$, hence the signal term has $N$ positive eigenvalues and $M-N$ zero eigenvalues if $M>N$.  There are $ M - N $ eigenvectors $\lambda_i$ such that $ AR_{SS}A^H\lambda_i = 0 $, also it satisfy $A^H \lambda_i=0$, finally we can find $\theta_i$ by $\lambda_i^H a(\theta) = 0$.

To sum up, the steps of MUSIC are followed by:
\begin{enumerate}
    \item Calculate $R_{XX}$,
    \item perform eigenvalue decomposition on $R_{XX}$,
    \item sort eigenvectors according to their eigenvalues in descent order,
    \item select last $M-N$ eigenvectors $\lambda_i$,
    \item calculate noise space matrix $ a^h(\theta_i) U_n U_n^H a(\theta_i)$
    \item plot $p(\theta)$ and find the peak.
\end{enumerate}

\begin{figure}[h]
\centering
    \includegraphics[width=1.0\linewidth]{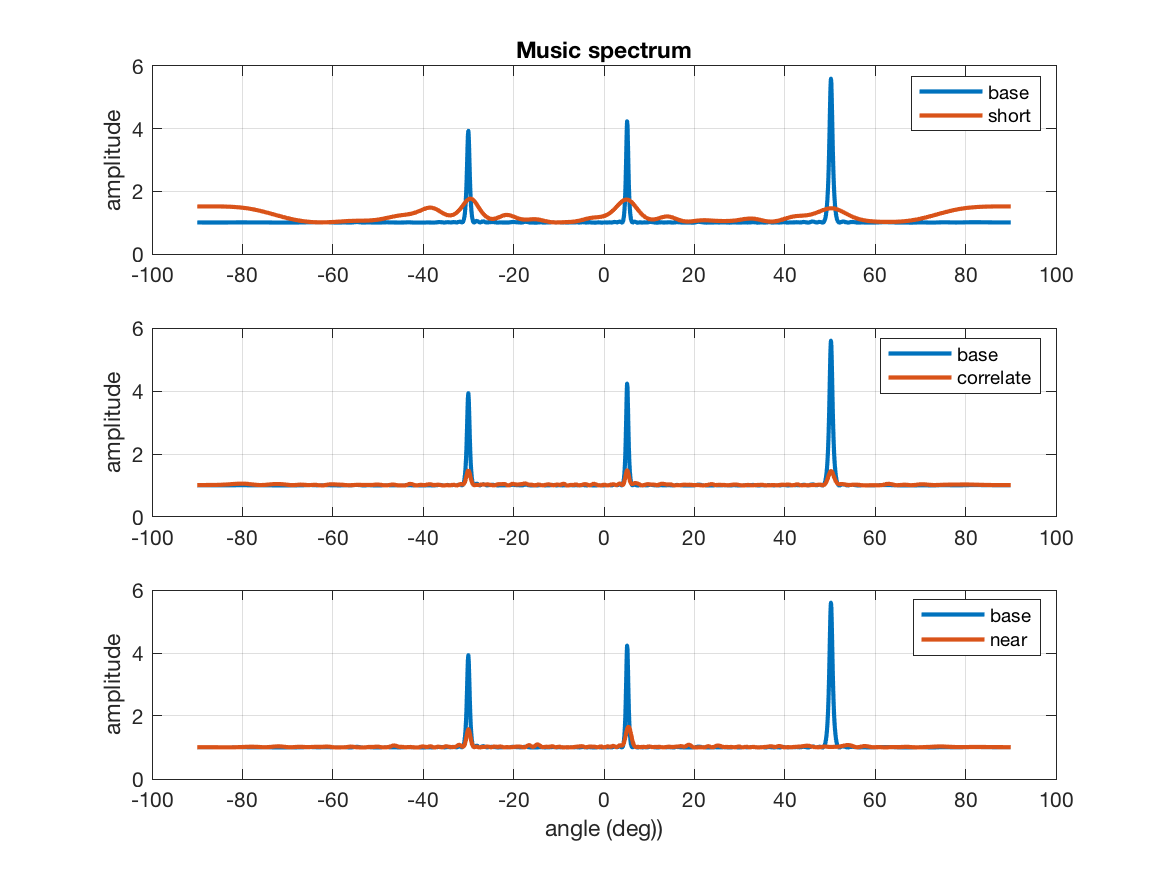}
    \caption{Disadvantage of MUSIC}
    \label{fig:music_spec}
\end{figure}

\begin{figure}[h]
\centering
    \includegraphics[width=0.6\linewidth]{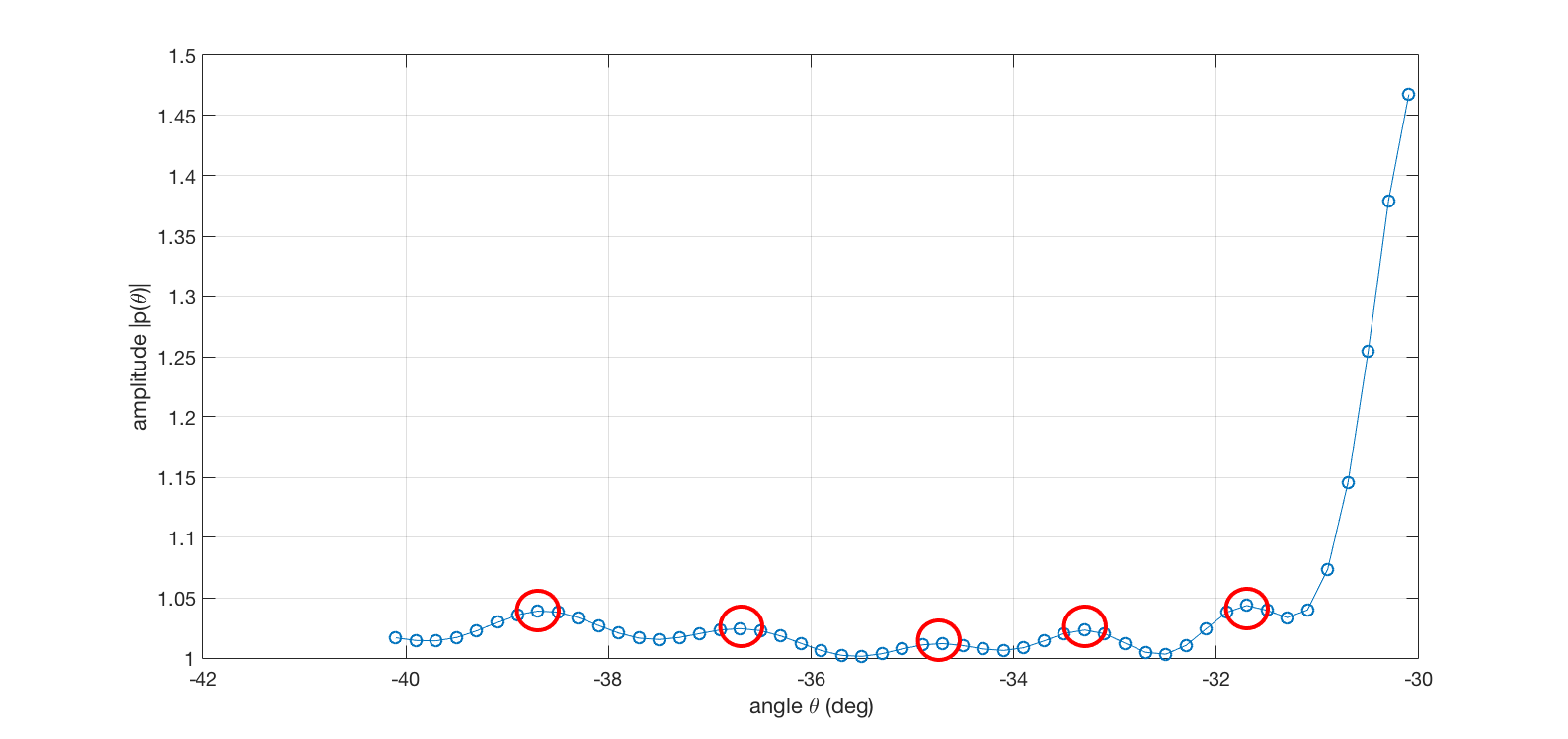}
    \caption{Illustration of finding local maximum point}
    \label{fig:findPeak}
\end{figure}

However, MUSIC does not specify how to find the peak, which in fact include large computation, especially for multiple sources. In our case, we sweep over all points, and find all turning points  i.e. local maximum, that meet $p(\theta_{i})> p(\theta_{i-1}),  p(\theta_i)> p(\theta_{i+1})$, and then pick the largest $K$ values of them, then the index $\theta$ is the estimator, as shown in Fig.~\ref{fig:findPeak}.

Meanwhile, as shown in Fig.~\ref{fig:music_spec} it also suffers from \textit{unresolvable issue}, which means two peaks may merge into one when two angles are near to each other. It also expects the source signal to be uncorrelated, otherwise it may fail to find the right peak.

\subsection{AoA estimation by neural networks}
 
The neural network could compensate for the drawback of MUSIC. On the view from the neural network, the problem is aimed to estimate the location out of raw data,
\begin{equation}
    (\hat{\theta_1},\hat{\theta_2}) = f_{NN}(R_{XX}),
\end{equation}
and here $f_{NN}(.)$ is the  approximate function given by neural network.

\subsubsection{Dataset synthesis}
Here the dataset is synthesis at local machine rather than field test.
These variables are randomly generated rather than step wised value, which would be a constraint to the minimum step, or resolution issue. As our demonstrate case is estimate azimuth $\theta_1$ and $\theta_2$, namely \textit{interested labels}, we randomly iterate $\theta_1$, $\theta_2$ in range $[-\pi/2,\pi/2]$ for 400 times, as well as white Gaussian noise at SNR 20dB, while keep other parameters fixed. Unlike the uniformly sweeping over the range, here there are randomly at no resolution without minimum step, and it turns out to work better.  Please notice that there is no guarantee to estimate beyond interesting angles. For example, if $\theta_1 \in [0,\pi/2]$ when generate training data, it is unlikely it will predict well with testing input where $\theta_1 \in [-\pi/2,0]$. Neither it would work when generating training data with source single carrier frequency at $f_1$, while testing data is from different carrier frequency $f_2$.

\subsubsection{Covariance matrix}

The single entry of raw channel estimation is matrix $A$ is of size $N\times M$, where $N$ for the number of snapshots and $M$ for the number of antennas. However, the size is too large as the input and over cost the computation. Hinted by MUSIC, we propose the \textit{covariance matrix} $Z$ 
\begin{equation}
    R_{XX} = A\times A',
\label{equ:cor}
\end{equation}
that could highly reduce the size while maintaining the information, and the size of  $R_{XX}$  is $N\times N$ in this way. Since  $R_{XX}$  is symmetric, only take the upper triangle part is enough. Also real and image needs to be extracted due to neural network deal only with real value. Hence, the size of input turn to be $M(M+1)$ now, which much smaller of the original size. 

To sum up, the data pre-processing is shown below.
\begin{enumerate}
  \item Calculates correlated matrix $R_{XX}$ as above,
  \item extract the upper triangle of the correlated matrix and convert into a vector,
  \item extract real and imaginary part of the vector and combine a new vector,
  \item normalize the vector, take it as the input.
\end{enumerate}

\subsubsection{Overfitting prevention}
 Overfitting refers to when a function is too closely fit a limited set of data points while too far for the other. In reinforcement learning, overfitting may happen as the function form by existing weight fit training dataset very well while poorly with testing data. Here we list some primary technic to prevent overfitting.

\begin{itemize}
    \item  \textit{Input regularization}, the element of the input value may get different scales, sometimes too large. Recent research \cite{Goodfellow-et-al-2016} show simple regular input to the same scale could help to avoid overfitting.
    \item \textit{Drop out}, by default ANN refer to fully connected between each layer. However, the study \cite{srivastava2014dropout} shows a layer may rely too much on a few of its inputs, make these at a larger value while others reach zeros. To this end, drop out method can randomly drop a small part of the input to break the high reliance. In code implementation, it is added as a kind of hidden layer, with all weights of value 0 or 1, which mean on or off to select which neuron is selected into connecting the next layer. The drop out rate is set as 0.10 in the evaluation.
    \item \textit{Weight decay}, similar to the over-relying issue, weight decay \cite{hanson1989comparing} is a regularization term that penalizes big weights. The weight decay loss is also added up to prediction loss. When the weight decay coefficient is big, the penalty for big weights is also big, when it is small weights can freely grow. 
    \item \textit{Batch normalization} \cite{ioffe2015batch}, during the training stage of networks, as the parameters of the preceding layers change, the distribution of inputs to the current layer changes accordingly, such that the current layer needs to constantly readjust to new distributions, this issue is named internal covariate shift. Recent study \cite{ioffe2015batch} show batch normalization allows each layer of a network to learn by itself a little bit more independently of other layers. It normalizes the output of a previous activation layer by subtracting the batch mean and dividing by the batch standard deviation.
    
    \begin{figure}[h]
\centering
    \includegraphics[width=.6\linewidth]{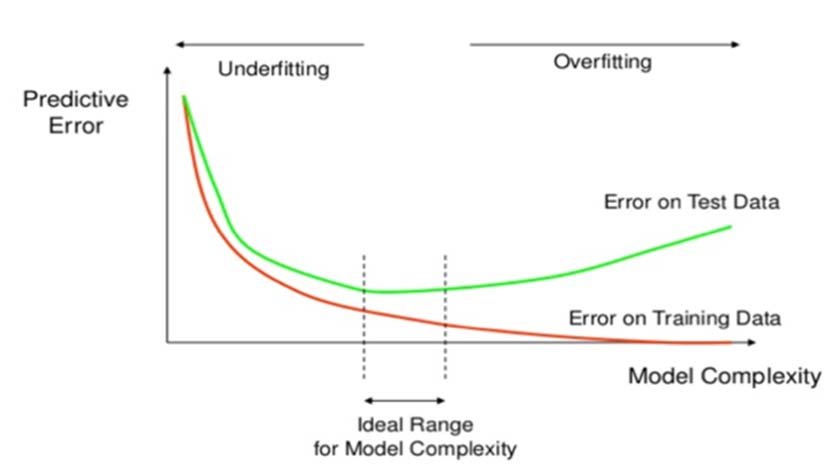}
    \caption{Early stop}
    \label{fig:early}
\end{figure}

    \item  \textit{Early stop} \cite{Goodfellow-et-al-2016}, this technical split a validate data set that never trained, but use for validation at each epoch. In the beginning, the validate loss and training loss both decrease rapidly, however, after some epoch, as shown in \ref{fig:early} training loss slows down while validate loss fluctuate or even goes up, and there is significant divergence between them. The early stop is executed to stop the deteriorate of overfitting.

\end{itemize}

\subsubsection{Radial based activation function}
The radial basis function (RBF) listed in ~\cite{park1991universal} is also verified here. This function is commonly interpreted as
\begin{equation}
    \phi(r) = e^{-(\sigma (r-c))^2},
\end{equation}
where $c$ is center vector determined by kNN method.
Its figure is shown Fig.~\ref{fig:activiation}, while the default option for activation function in \textit{relu}.

\subsubsection{Bayesian neural network}
A Bayesian neural network ~\cite{blundell2015weight} is a neural network with a prior distribution on its weights.
Consider a data set $\{(x_n, y_n)\}$, where each data point comprises of features ${x}_n\in\mathbb{R}^D$ and output $y_n\in \mathbb{R}$. Define the likelihood for each data point as
\begin{equation}
    p(y_n \mid \mathbf{w}, \mathbf{x}_n, \sigma^2) = \text{Normal}(y_n \mid \mathrm{NN}(\mathbf{x}_n;\mathbf{w}), \sigma^2),
\end{equation}
where NN is a neural network whose weights and biases form the latent variables $\mathbf{w}$. 
Define the prior on the weights and biases $\mathbf{w}$ to be the standard normal
\begin{equation}
 p(\mathbf{w}) = \text{Normal}(\mathbf{w} \mid \mathbf{0}, \mathbf{I}).
\end{equation}

\begin{figure}[h]
\centering
    \includegraphics[width=.6\linewidth]{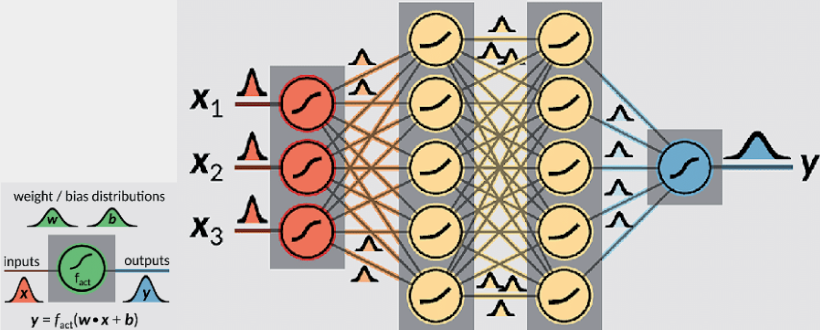}
    \caption{Bayesian Neural Network}
    \label{fig:bnn}
\end{figure}
BNN is claims to perform with datas coming from specific model and require smaller size datasize, which make it worthy in our problem.

\subsubsection{Multiple output schemes}
\begin{figure}[h]
\centering
    \includegraphics[width=1.0\linewidth]{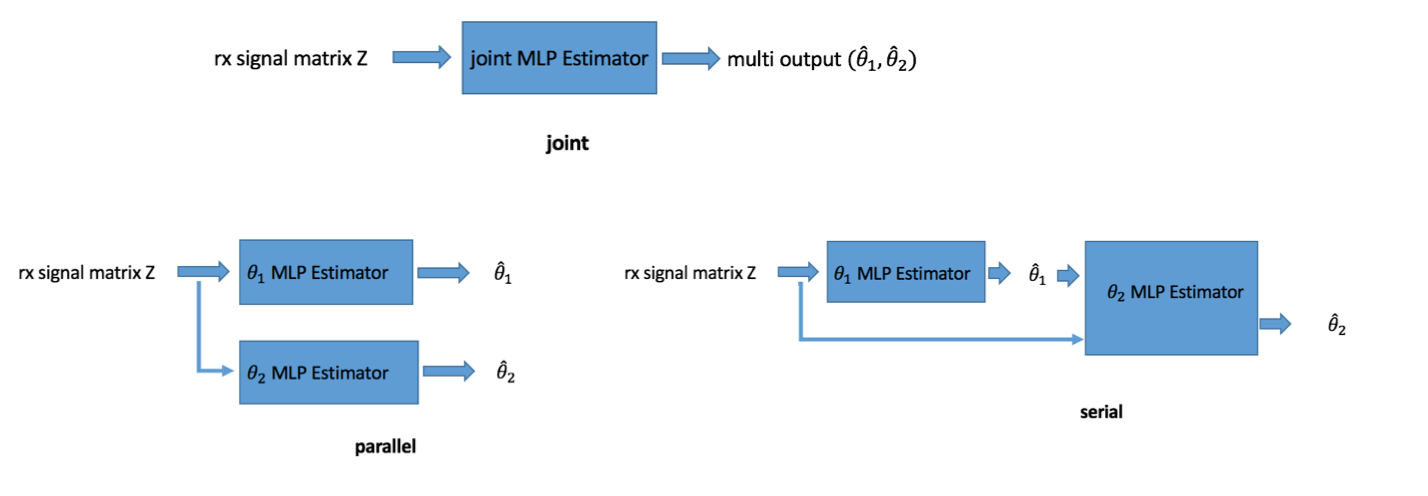}
    \caption{Multiple output scheme}
    \label{fig:multi}
\end{figure}

Since our neural network work not only for one angle estimation but also work for several angles estimation. 
\begin{enumerate}
    \item joint, one straightforward way is output all expect angles $(\hat{\theta_1}, \hat{\theta_1}, ..., \hat{\theta_M})$ at one time, as named as \textit{joint} scheme.
    \item parallell, a dedicate neural network could be used to estimate one angle only, regardless of other angles exist or not. So there needs to be $M$ neural network for estimate $M$ angles, and this scheme is named \textit{parallel}.
    \item  serial, we could take the first output $\hat{\theta_1}$ as the input to estimate next angle  $\hat{\theta_2}$, and then take exiting angles $\hat{\theta_1}, \hat{\theta_2} $ to estimate the third one, and repeat till getting all interesting angles, and it is named \textit{serial} scheme. 
\end{enumerate}
 Notice serial differs from parallel for it gets more and more number of input at later block. All of them is illustrated in Fig.~\ref{fig:multi} More output predicts also mean larger network size and more training time.

\subsection{Peformance evaluation}
Our preliminary work is written in Python, with help mainly from \textit{tensorflow} and \textit{keras} library. These codes are running on the remote machine of Advanced Research Center (ARC) at Virginia Tech.

\begin{table}[h]
\caption{ULA Setting}
\label{table:ula}
\begin{center}
\begin{tabular}{|c|c|c|}
\hline
Notation & Setting & Value\\
\hline
$d$ & spacing (meter) & 0.10\\
\hline
$L$ & snapshot length  & 4,000\\
\hline
$M$ & number of source & 3\\
\hline
$N$ & number of antennas & 6\\
\hline
$f_c$ & source signal carrier frequency (Hz) & $10^6$\\
\hline
$P_c$ & source signal power (Watt) & $1.0$\\
\hline

\end{tabular}
\end{center}
\end{table}

Some other detailed setting is shown in Table.~\ref{table:ula}. Notice the source single $P_c$, $f_c$ does not necessarily to be the same. To prevent overfitting, we here adopt $K$-fold $cross-validate$ for neural network, which means each time $1/K$ slice of training data is cut as validation data while remaining taken into training, and this cut-validate is repeated for $K$ times, and a set of $K$ training weighted as saved. Finally the average of these weights $\bar{\textbf{}w}$ is chosen as the final weighted.
\begin{table}[h]
\caption{Neural Network Setting}
\label{table:settings}
\begin{center}
\begin{tabular}{|c|c|}
\hline
Setting & Value\\
\hline
\hline
Test ratio & 0.10\\
\hline
Number of testing samples & 2000\\
\hline
Batch Size & 64\\
\hline
Hidden layers & [25 25 25]\\
\hline
Drop out ratio & 0.10\\
\hline
Batch normalization L2 regularizers & 0.001\\
\hline
Optimizer & \textit{Adam}\\
\hline
Loss function & \textit{mse}\\
\hline
Drop out ratio & 0.10\\
\hline

\end{tabular}
\end{center}
\end{table}

\subsubsection{Neural network options}

\begin{figure}[h]
\centering
    \includegraphics[width=.6\linewidth]{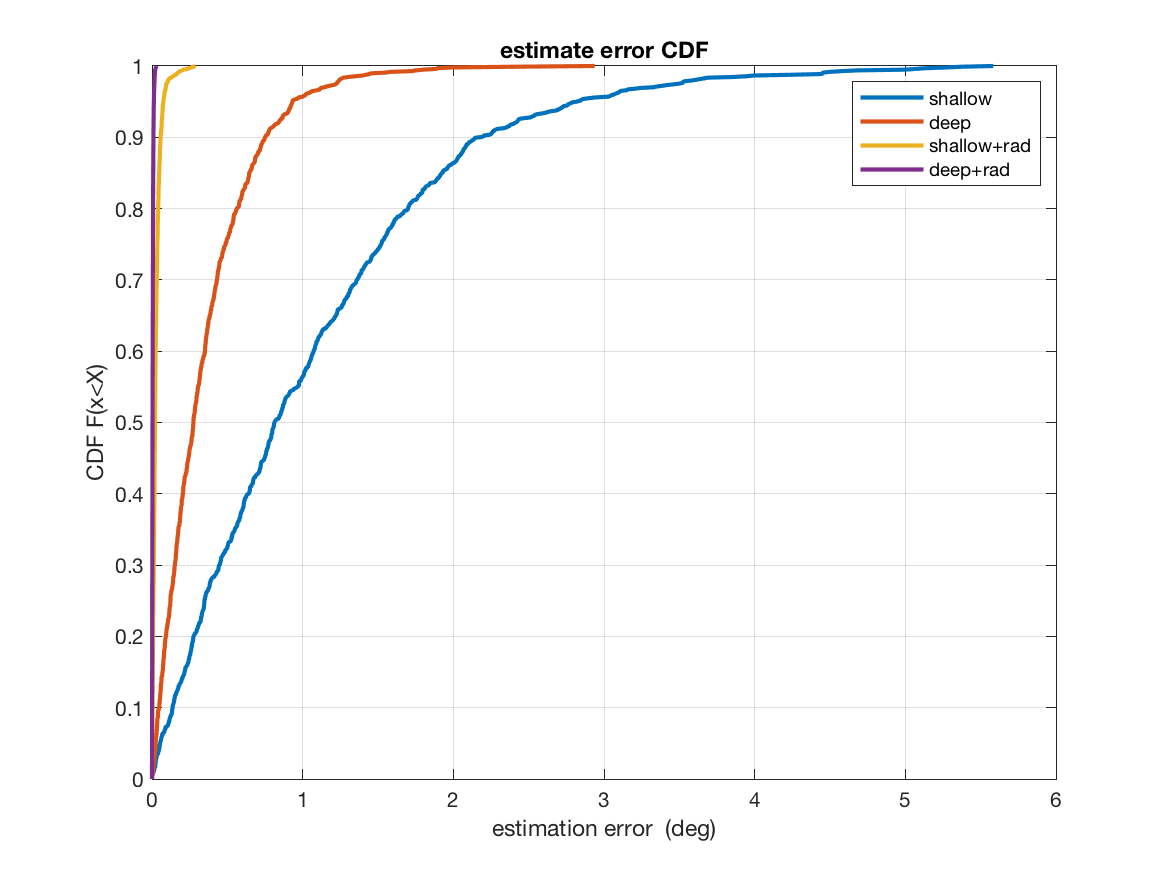}
    \caption{Error CDF of RBF}
    \label{fig:cdf_rad}
\end{figure}

\begin{figure}[h]
\centering
    \includegraphics[width=.6\linewidth]{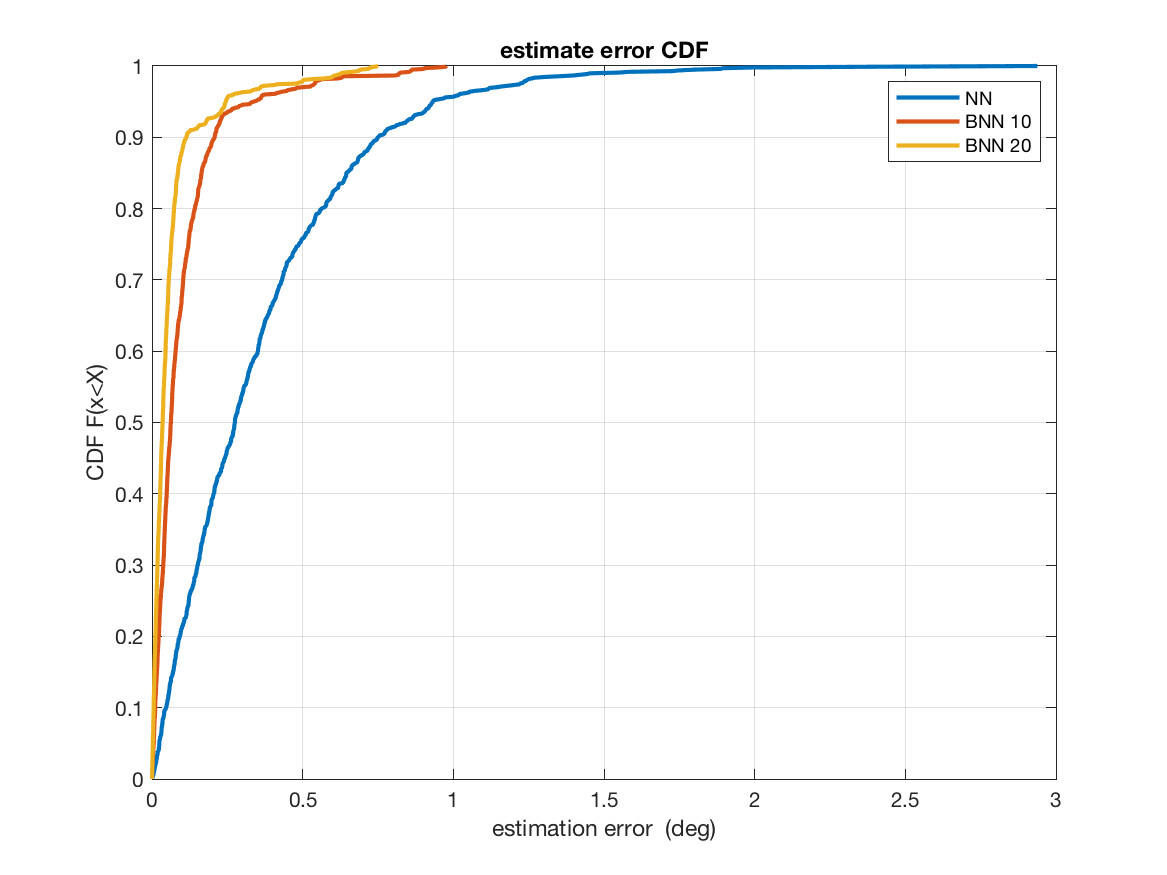}
    \caption{Error CDF of BNN}
    \label{fig:cdf_bnn}
\end{figure}

Fig.~\ref{fig:cdf_rad} shows deeper network of size $[25,25,25]$ can reduce error better than shallow network $[75]$, and RBF can further assist the estimation.  Fig.~\ref{fig:cdf_bnn} shows BNN can get more precise estimation than default  Scaled Conjugate Algorithm (SCG), and longer epoch brings better result. There is not validate data and early stop scheme in BNN, so it requires a maximum epoch to end the computing. Notice that since RBF take much longer time per epoch than GPU-assisted relu, and GPU support SCG rather than BNN, we still adopt the later as the default. 

\subsubsection{Signal noise Ratio}
\begin{figure}[h]
\centering
    \includegraphics[width=.6\linewidth]{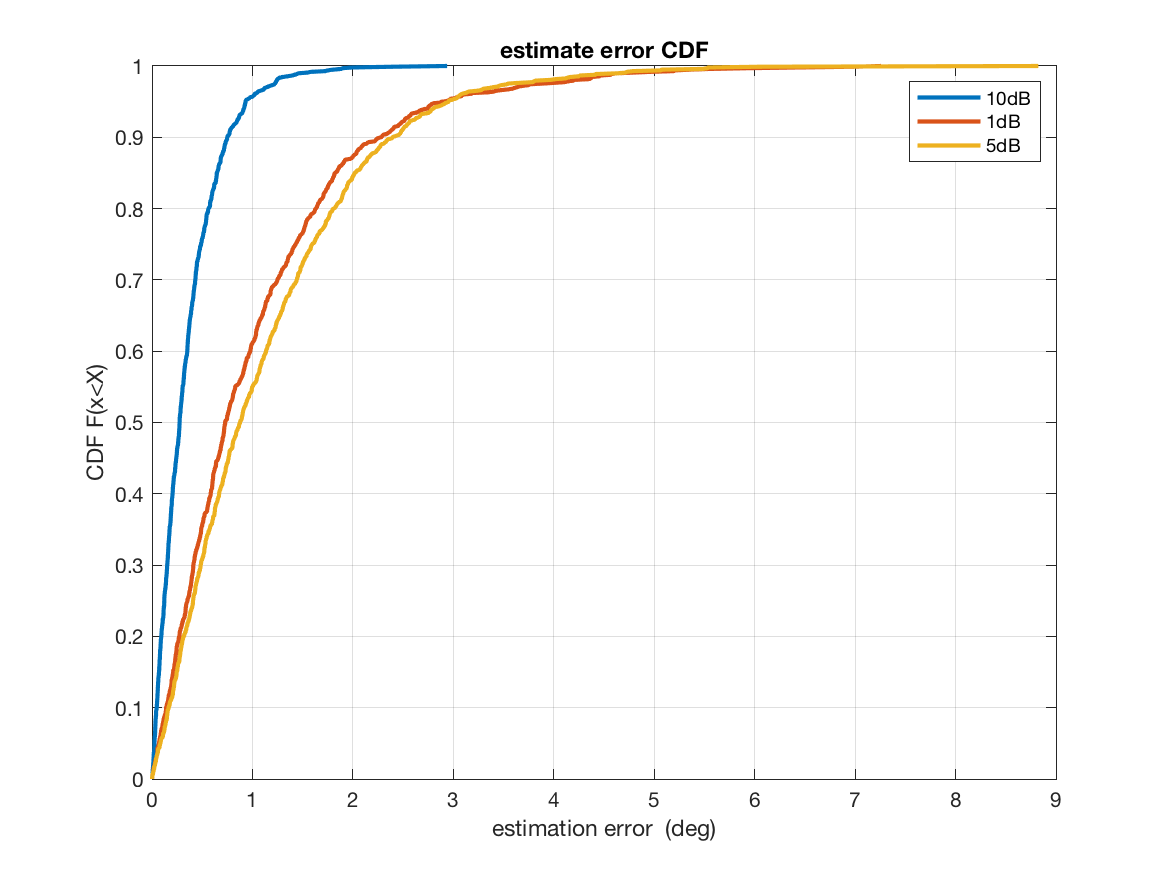}
    \caption{Error CDF of different SNR}
    \label{fig:cdf_SNR}
\end{figure}

As the most common interference in wireless communication. We also evaluation neural network performance over SNR, it turns out the neural network is quite robust to white Gaussian noise as shown in Fig.~\ref{fig:cdf_SNR}. Additionally, once if the SNR could be measured before doing the estimation, this metric could also add up to the inputs.

\subsubsection{Nonuniform antenna spacing}
\begin{figure}[h]
\centering
    \includegraphics[width=.6\linewidth]{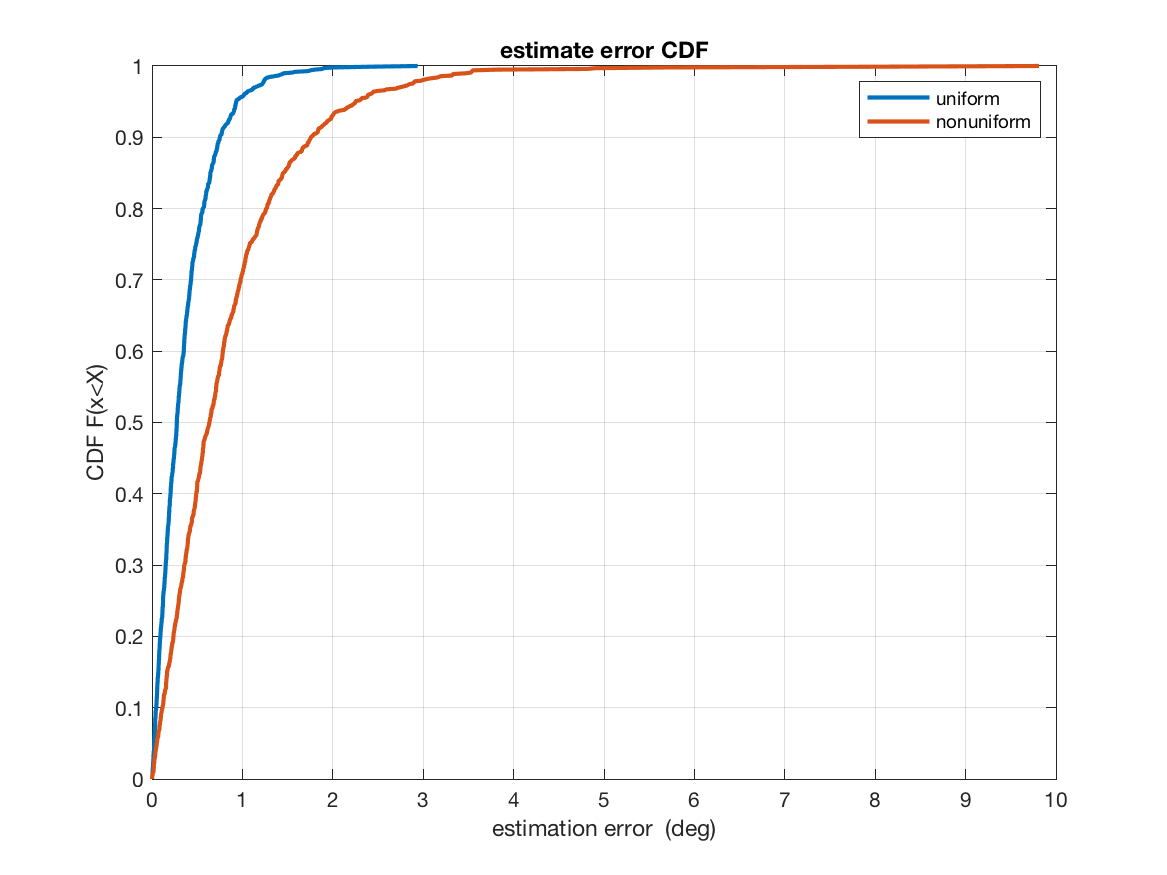}
    \caption{Error CDF of nonuniform spacing}
    \label{fig:cdf_spacing}
\end{figure}

Practical issue such as imperfect geometry and clutter will also bring error to AoA estimation. The classic method such as MUSIC can also deal with it \cite{liu1998azimuth}. We here consider the nonuniform spacing case, which should be theoretical same while may not in reality. The signal model of non uniform linear array follow the equation
\begin{equation}
x_m(t) = \sum_{n=1}^N s_n(t)e^{j2\pi d_m cos(\theta_n)/\lambda}+n_m(t)
\end{equation}, and $d_m = (m-1)d + \epsilon$, where   $\epsilon$ is the tiny bias. As shown in Fig.~\ref{fig:cdf_spacing}, neural network is robust to such basis and still work well under this case.

\subsubsection{Comparison to classic methods}

\begin{figure}[h]
\centering
    \includegraphics[width=.6\linewidth]{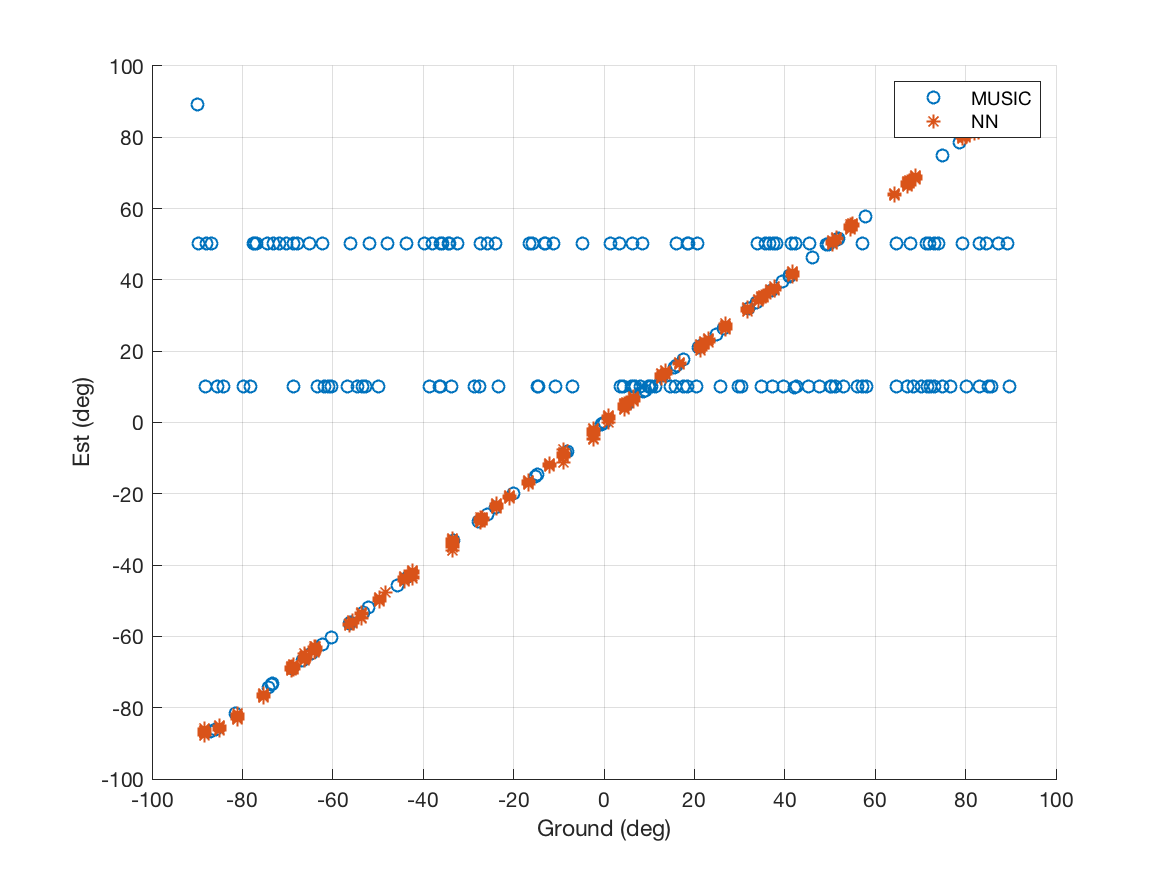}
    \caption{Scatter plot of NN and MUSIC }
    \label{fig:scatter}
\end{figure}

\begin{figure}[h]
\centering
    \includegraphics[width=.6\linewidth]{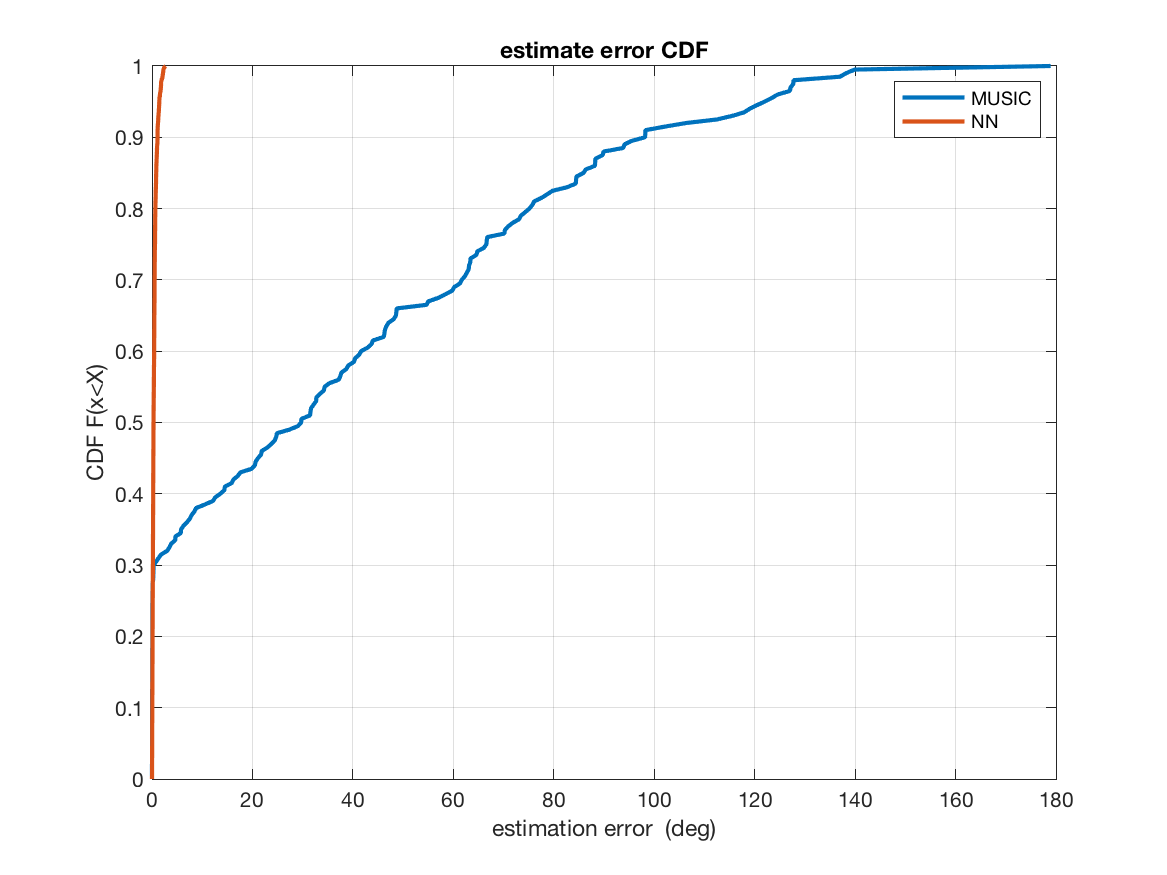}
    \caption{Error CDF of NN and MUSIC}
    \label{fig:cdf_music}
\end{figure}
The comparison between Neural Network and MUSIC is shown in Fig.~\ref{fig:cdf_music}, as we can see neural network significantly perform better than MUSIC in our case, given the same dataset. Here the dataset resolution is set at 0.01 degree, and MUSIC search resolution is 0.01. As shown in Fig.~\ref{fig:scatter}, the major problem of MUSIC is the unresolvable issue, it would fail once if two angles are near to each other. However, neural network is not constraint to such issue.

\subsection{Conclusion}
Based on the previous evaluation, we can see the neural network can be a potential solution for AoA estimation for its compatible performance with MUSIC. Meanwhile, the RBF and BNN proof to benefit the estimate accuracy, and our estimator is robust to noise and imperfect geometry. However, the neural network may be constrained to the training dataset, meaning there is no guarantee once predict data never shown around the training data.

\subsection{Future work}
Our current work shows neural network can well estimate DoA at high precision over MUSIC gave proper dataset. Moreover, there are some more aspects we can continue our work in the further.
\begin{enumerate}
    \item Evaluate how well NN can distinguish the adjacent angles.
    \item Evaluate the performance versus other classic AoA methods such as MVDR and ESPRIT, and compare to other classic machine learning methods like SVM and kNN, as well as evaluate CNN-based neural network estimator,
    \item For the diversity of input signal, we can evaluate other types of geometry of array, like a uniform circular array (UCA), also evaluate its performance with some other modulation scheme like digital spectrum spreading, along with evaluation under interference, like narrowband interference.
    \item For multiple outputs, we can scale up the multiple AoA estimation evaluation.
\end{enumerate}

\newpage
\bibliographystyle{IEEEtran}
\bibliography{eqbib}
\footnotesize{
}

\end{document}